\documentclass[a4paper]{article}

\usepackage{INTERSPEECH2021}
\usepackage{cite}
\usepackage{bm}
\usepackage{multirow}

\usepackage{mathrsfs}
\usepackage{amsmath}
\usepackage{url}

\title{Improving Channel Decorrelation for Multi-Channel Target Speech Extraction}
\name{Jiangyu Han$^{1,2}$, Wei Rao$^{2}$, Yannan Wang$^{2}$, Yanhua Long$^{1}$\thanks{Work was done when Jiangyu Han was an intern at Tencent Ethereal Audio Lab. Yanhua Long is the corresponding author. The work is supported by the National Natural Science Foundation of China (Grant No.62071302).}}

\address{
  $^1$Shanghai Engineering Research Center of Intelligent Education and Bigdata, \\
Shanghai Normal University, Shanghai, China\\
  $^2$Tencent Ethereal Audio Lab, Tencent Corporation, Shenzhen, China}
\email{jyhan03@163.com, \{ellenwrao,yannanwang\}@tencent.com, yanhua@shnu.edu.cn}

\begin{document}

\maketitle
\begin{abstract}
\label{sec:abs}

Target speech extraction has attracted widespread attention.
When microphone arrays are available, the additional
spatial information can be helpful in extracting the target speech.
We have recently proposed a channel decorrelation (CD) mechanism to
extract the inter-channel differential information to enhance the reference
channel encoder representation. Although the proposed mechanism has shown promising
results for extracting the target speech from mixtures,
the extraction performance is still limited by the nature of the original
decorrelation theory. In this paper, we propose two methods to
broaden the horizon of the original channel decorrelation,
by replacing the original softmax-based inter-channel similarity between
encoder representations, using an unrolled probability and a normalized
cosine-based similarity at the dimensional-level. Moreover,
new combination strategies of the CD-based spatial information
and target speaker adaptation of parallel encoder outputs are also
investigated. Experiments on the reverberant WSJ0 2-mix show that the improved
CD can result in more discriminative differential information
and the new adaptation strategy is also very effective
to improve the target speech extraction.

\end{abstract}


\section{Introduction}
\label{sec:intro}

Target speech extraction (TSE) aims to extract only a target
signal from the mixed speech \cite{tse}. Generally, most TSE approaches
require additional target speaker clues to drive the network
towards extracting the speech of that speaker.
Many previous works have focused on the TSE tasks \cite{tse_related1,
tse_related2,tse_related3,tse_related4,tse_related5}. Although these works
have achieved great success on close-talk target speech extraction tasks,
the performance of far-field speech extraction is still
far from satisfactory due to the reverberation, noise, etc.
In that case, the additional multi-channel spatial information
usually can be used to improve the separation ability.

Many contributions have been paid to exploit the spatial information
between multi-channel recordings for TSE. For example, the direction aware
SpeakerBeam \cite{direct} combines an attention mechanism with
beamforming to enhance the signal of the target direction; the neural spatial filter
\cite{neural} uses the directional information of the target speaker to
extract the corresponding speech; the time-domain SpeakerBeam (TD-SpeakerBeam)
\cite{tsb} incorporates the inter-microphone phase difference (IPD) \cite{ipd}
as additional input features to further improve
the speaker discrimination capability. All of these
approaches showed promising results, which indicates that the multi-channel information
can provide an alternative guider to discriminate the target speaker better.
Actually, one of the keys of TSE is still speech separation that
refers to extract all overlapping speech sources from the mixed speech \cite{ss}.
In order to further improve the separation ability, many strategies for exploiting
the multi-channel information have been recently proposed, such as normalized crosscorrelation (NCC) \cite{ncc}, transform-average-concatenate (TAC) \cite{tac}, and inter-channel convolution difference
(ICD) \cite{icd}, etc. Therefore, how to effectively exploit the multi-channel spatial
information for TSE is crucial.

Considering that the mixed speech in an $N$-dimensional embedding space
can be decomposed into $N$ representations at the dimensional-level. Given
multi-channel input mixtures, each dimensional-level representation
among different microphone channels not only has high correlations, but also has some
differentiations, both of these correlations and differentiations
characterize the inter and intra speaker properties in the mixture.
Our previous work in \cite{cd} proposed a channel decorrelation (CD) mechanism
with parallel encoder target speaker adaptation to leverage these information
for improving TSE systems.
This decorrelation is performed on each dimension of all the multi-channel
encoder representations of input mixtures, it is used to
extract the inter-channel differential spatial information to
learn difference between individual source signals of input mixture.
Results in \cite{cd} have already shown that our original CD
significantly improved the TSE performance over IPD features,
however, the performance gains over the parallel encoder architecture \cite{para}
are still limited.

Therefore, in this study, we revisit the principle of our original channel
decorrelation algorithm, two improvements are proposed to replace the original
softmax-based inter-channel similarity calculation, one is using an unrolled
probability and the other is using a normalized cosine-based similarity.
Both of them are performed at the dimensional-level of encoder representations,
and they aim to expand the dynamic range of the inter-channel differential probability
to capture the fine-grained representation
for a better spatial information utilization. Moreover, new combination strategies of the improved CD
and target speaker adaptation of parallel encoder outputs are also
investigated. All of experiments are performed on the publicly available multi-channel
reverberant WSJ0 2-mix dataset. Results show that both of our improved channel
decorrelation and the new adaptation strategy are very effective to improve
the target speech extraction.

\section{Related Works}
\label{sec:rw}

\subsection{Conv-TasNet}

The Conv-TasNet is a time-domain speech separation technique proposed in \cite{tasnet}. It has become the mainstream speech separation approach
because of its competitive performance over most of time-frequency domain
speech separation algorithms. This separation structure has attracted widespread attention and
further improved in many recent works \cite{tasnet_relate1,tasnet_relate2,tasnet_relate3,tasnet_relate4}.
Conv-TasNet consists of three parts: an encoder (1d convolution layer),
a mask estimator (several convolution blocks), and a decoder (1d deconvolution layer).
The waveform mixture is first encoded by the encoder and then is fed into the temporal
convolutional network (TCN) \cite{tcn} based mask estimator to
estimate a multiplicative masking function for each source. Finally, the source
waveforms are then reconstructed by transforming the masked encoder representations
using the decoder. More details of Conv-TasNet can be found in \cite{tasnet}.

\subsection{TD-SpeakerBeam}

TD-SpeakerBeam is a Conv-TasNet based target speech extraction approach
that has been recently proposed in \cite{tsb}. As shown in Fig.\ref{fig:overviews} (a),
it introduces an auxiliary network with an encoder and a convolution
block to transform a pre-saved target speaker enrollment waveform to
a target speaker embedding vector. This vector is then used in a
scaling adaptation layer \cite{sa} to drive the network to extract the
corresponding speech. The encoder and convolution block in
the auxiliary network have the same structure as they used in Conv-TasNet.
The speaker adaptation is performed between the first and second convolution blocks
of the mask estimator. And in the adaptation layer,
the mixture representation matrix is re-weighted by an element-wise multiplication
with the repeated target speaker embedding vectors.
Furthermore, by concatenating the IPD features after
the adaptation layer, the performance of TD-SpeakerBeam can be further improved.

In Fig.\ref{fig:overviews} (a), $\mathbf{y_1}$,
$\mathbf{\hat{x}}^s$, $\mathbf{a}^s$, and $\mathbf{e}^s$ are the mixture waveform
of the first channel, the extracted target speech waveform, the enrollment utterance
of the target speaker, and the target speaker embedding vector respectively.
We use this IPD configuration as one of our baseline.

\section{Channel Decorrelation}
\label{sec:cd}

\subsection{Original CD}
\label{subsec:ocd}

\begin{figure}[t]
  \centering
  \includegraphics[width=8.2cm]{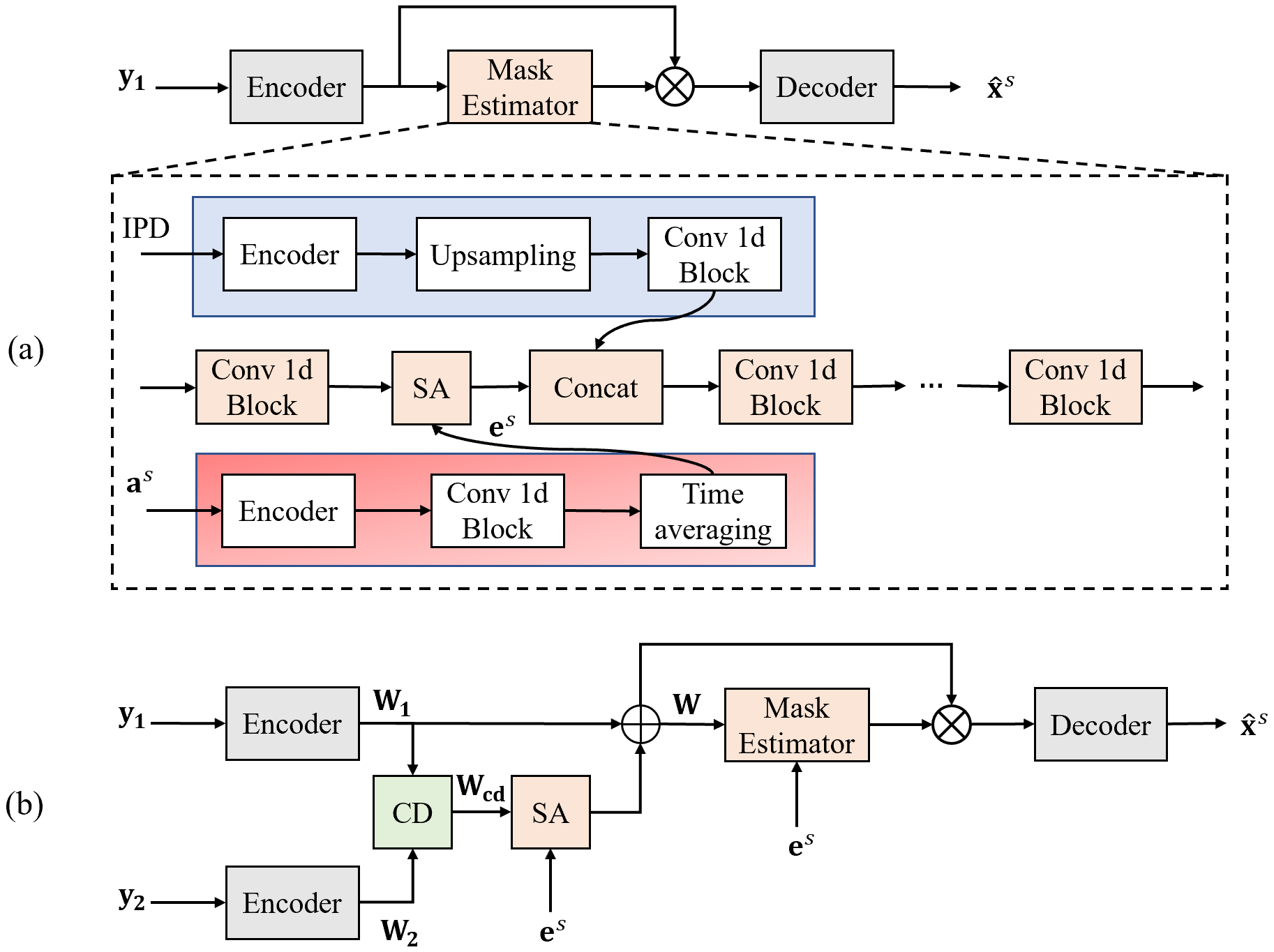}
  \caption{Conv-TasNet based structures that incorporated with IPD features and channel decorrelation (CD) mechanism for the target speech extraction task. All network modules are the same as Conv-TasNet. ``SA" indicates scaling adaptation.}
  \label{fig:overviews}
\end{figure}

Assuming there is a $N$-dimensional embedding space, each channel of the
multi-channel mixture waveform can be encoded into a $N \times T$ embedding matrix,
where $T$ is the length of that matrix. Each dimensional-level vector of the
mixture embedding can be seen as one type of fine-grained representation that
corresponds to different local traits of the whole input mixture.
Our channel decorrelation is proposed to construct such a latent embedding space
to learn dimensional-level discriminate feature representations between
the encoded representations of multi-channel input mixtures, by
taking the advantage of spatial configuration of multi-channel microphone-arrays.
Taking two channels for example, the CD process is shown in Fig.\ref{fig:overviews} (b).
It accepts two encoded mixture representations $\mathbf{W_1}$ and $\mathbf{W_2}$,
and then outputs the differential spatial information $\mathbf{W_{cd}}$ between
two channels. The specific calculation of $\mathbf{W_{cd}}$ is as follows:

First, compute the cosine correlation between $\mathbf{W_1}$
and $\mathbf{W_2}$ in each corresponding dimension (row).
Note that $\mathbf{W_1}$ and  $\mathbf{W_2}$ are two matrices, i.e.,
\begin{equation}
\mathbf{W_i} = {[\mathbf{w_i}^1, \mathbf{w_i}^2, ..., \mathbf{w_i}^N]}^\mathbf{T}, i = 1, 2
\end{equation}
where $\mathbf{W_i} \in \mathbb{R}^{N \times T}$ is the input
of the $i$ th channel of CD mechanism, $N$ is the output dimension
of convolutional encoder, $T$ is the length of convolutional encoder output.
$\mathbf{w_i}^j \in \mathbb{R}^{1 \times T}, j = 1, 2, ..., N$,
is the $j$ th dimension vector of the $i$ th channel.
$\mathbf{T}$ is the operation of transpose.

The cosine correlation of $j$ th dimensional vector between
the first channel and the second channel is calculated as,
\begin{equation}
\phi_{1,2}^j = \frac{\langle \mathbf{w_1}^j, \mathbf{w_2}^j \rangle}{\lVert{\mathbf{w_1}^j}\rVert_2 \lVert{\mathbf{w_2}^j}\rVert_2},  j = 1, 2, ..., N
\end{equation}
where $\langle \cdot \rangle$ is the inner product of two vectors,
$\lVert\cdot\rVert_2$ represents the Euclidean norm.
The vectors involved in the operation are normalized to zero mean prior to the calculation.

Then, calculate the cosine correlation between the vectors of each
dimension of $\mathbf{W_1}$ and  $\mathbf{W_2}$ in turn, and concatenate them to
get a similarity vector $\bm{\phi_{1,2}}$,
\begin{equation}
\bm{\phi_{1,2}} = [\phi_{1,2}^1, \phi_{1,2}^2, ..., \phi_{1,2}^N]^\mathbf{T}
\end{equation}
where $\bm{\phi_{1,2}} \in \mathbb{R}^{N \times 1}$ represents the
similarity of two encoded mixture representations
in each dimension in a latent space.

Next, introduce an auxiliary vector $\mathbf{a}$ with the same size
as $\bm{\phi_{1,2}}$ and values are all 1, i.e.,
\begin{equation}
\mathbf{a} = [1,1,...,1]^\mathbf{T}
\end{equation}
where $\mathbf{a} \in \mathbb{R}^{N \times 1}$ can be regarded as the
cosine correlation between first channel $\mathbf{W_1}$ itself,
which represents the reference vector of $\bm{\phi_{1,2}}$.
Then, a softmax operation is used to
calculate the probability between each corresponding
dimensional element in $\bm{\phi_{1,2}}$ and $\mathbf{a}$ as $p_{1,2}^j$ to
get a similarity probability vector $\mathbf{p_{1,2}} \in \mathbb{R}^{N \times 1}$ as,
\begin{equation}
\begin{split}
p_{1,2}^j &= \frac{e^{\phi_{1,2}^j}}{e + e^{\phi_{1,2}^j}}, j = 1, 2, ..., N \\
\mathbf{p_{1,2}} &= [p_{1,2}^1, p_{1,2}^2, ..., p_{1,2}^N]^\mathbf{T}
\label{eq:p12}
\end{split}
\end{equation}

Next, subtract $\mathbf{p_{1,2}}$ from $\mathbf{a}$ to get
a vector $\mathbf{s_{1,2}} \in \mathbb{R}^{N \times 1}$
that represents differentiated scores between channels,
and repeat $\mathbf{s_{1,2}}$ to the same size
as $\mathbf{W_2}$ to get the differentiated score matrix $\mathbf{S_{1,2}} \in \mathbb{R}^{N \times T}$, i.e.,
\begin{equation}
\begin{split}
\mathbf{s_{1,2}} &= \mathbf{a - p_{1,2}} \\
\mathbf{S_{1,2}} &= [\mathbf{s_{1,2}}, \mathbf{s_{1,2}}, ..., \mathbf{s_{1,2}}]
\label{eq:s12}
\end{split}
\end{equation}

Finally, the differentiated spatial information $\mathbf{W_{cd}}$
between channels can be obtained by multiplying
$\mathbf{W_2}$ by $\mathbf{S_{1,2}}$ as
\begin{equation}
\mathbf{W_{cd}} = \mathbf{W_2} \odot \mathbf{S_{1,2}}
\end{equation}
where $\mathbf{W_{cd}} \in \mathbb{R}^{N \times T}$, $\odot$ denotes
element-wise multiplication. In addition, to better guide the network
towards extracting speech of target speaker, we also
perform the target speaker scaling adaptation on $\mathbf{W_{cd}}$
to exploit the target speaker-dependent spatial information, as
shown in Fig.\ref{fig:overviews} (b).

\subsection{Unrolled CD}
\label{subsec:ucd}

Note that in Equation (\ref{eq:p12}), the softmax-based inter-channel similarity
$p_{1,2}^j$ ranges from $\frac{1}{e^2+1}$ to $\frac{1}{2}$ or $0.12$ to $0.5$, which
results in a $\mathbf{s_{1,2}}\in[0.5, 0.88]$ differentiated score range.
It means that the original CD  tends to assign a bigger but coarse weight
on the inter-channel differential information than the correlations.
This is incompatible with our motivation. Intuitively, ideal inter-channel differential
information of the same utterance from different microphone channels
should be less but fine and discriminative enough.

In order to alleviate such a problem, we modify the Equation ($\ref{eq:p12}$) as
\begin{equation}
p_{1,2}^j = \frac{2e^{\phi_{1,2}^j}}{e + e^{\phi_{1,2}^j}}, j = 1, 2, ..., N
\label{eq:unfold}
\end{equation}
By doing so, we can achieve a $[0.24, 1]$ inter-channel similarity range and
the resulting differentiated score range becomes $[0, 0.76]$.
We call this modification as ``unrolled CD''.
By expanding the dynamic range of $\mathbf{s_{1,2}}$ distribution,
the unrolled CD may result in more fine and discriminative inter-channel
differential information.

\subsection{Cosine-based CD}
\label{subsec:ccd}

Another method to address the limitation of original CD that
mentioned in section \ref{subsec:ucd} is to directly normalize
the cosine-based inter-channel similarity to [0,1] using the conventional
cosine probability transformation $(1+\phi_{1,2}^j)/2$.
In this way, the resulting differentiated score is defined as
\begin{equation}
\mathbf{s_{1,2}} = \frac{\mathbf{a} - \bm{\phi_{1,2}}}{2}
\label{eq:cos}
\end{equation}
And the dynamic range of $\mathbf{s_{1,2}}$  becomes $[0,1]$.
We call this modification as ``cosine-based CD''.

\subsection{CD with Parallel Encoder Speaker Adaptation}
\label{sec:impadpat}

\begin{figure}[t]
  \centering
  \includegraphics[width=8.2cm]{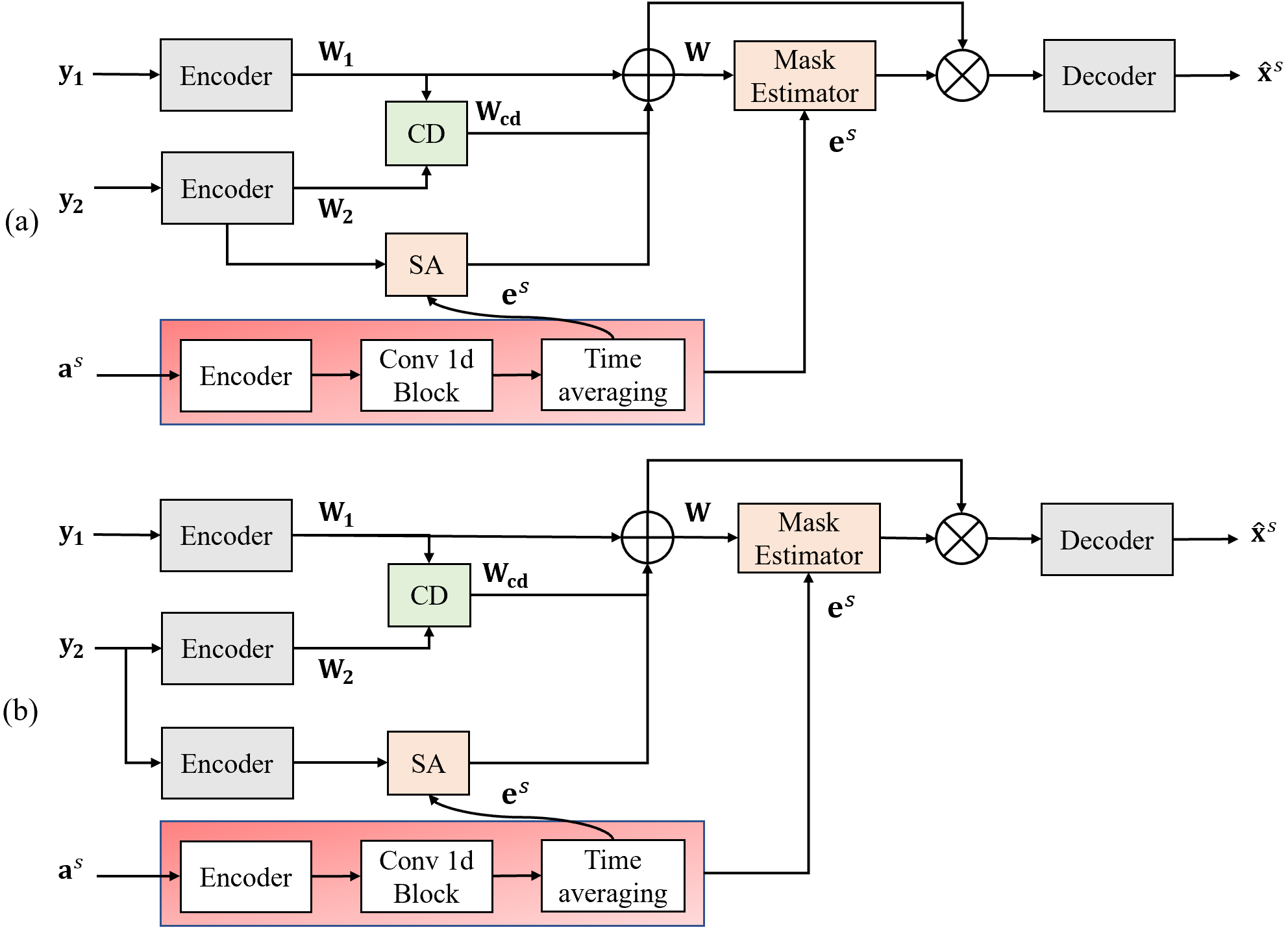}
  \caption{Different combination strategies of the CD-based spatial information
and target speaker adaptation of parallel encoder outputs.}
  \label{fig:cd_sa}
\end{figure}

Besides the improvements of the original CD algorithm, we also investigate
new combination strategies of the CD-based spatial information
and target speaker adaptation of parallel encoder outputs.
Subfigure (a) and (b) in Fig.\ref{fig:cd_sa} demonstrate two
new different combination strategies, in both of which the
output of CD only plays a role of effective inter-channel discriminative
spatial information as the IPD used in TD-SpeakerBeam.
These two proposals are motivated by the speculation that
the distribution of $\mathbf{W_{cd}}$ is very sparse.
Such a sparse distribution may not contain enough differential
spatial information to well characterize the target speaker identity.

\section{Experiments}
\label{sec:exp}

\subsection{Dataset}
\label{subsec:data}

Our experiments are performed on the multi-channel reverberant
WSJ0-2mix corpus \cite{wsj0}. Multi-channel recordings are generated by
convolving clean speech signals with room impulse responses simulated
with the image method for reverberation time of up to about 600ms \cite{tsb}.
The dataset consists of 8 channel recordings, but to have a fair comparison
with the state-of-the-art baselines, we also use only two channels in experiments.

We use the same way as in \cite{xu} to generate adaptation
utterances of the target speaker. The utterance is anechoic and
is selected randomly that different from the one in the mixture.
The size of training, validation, and test sets are 20k, 5k, and 3k utterances, respectively.

\subsection{Configurations}
\label{subsec:cfg}

All experiments are performed based on our previously released open source
of the original CD-based TSE in \cite{cd,cdcode}.
We still use the same hyper-parameters as our the original CD \cite{cd}.
However, it's worth noting that in this study, we only use
the scale-invariant signal-to-distortion ratio (SI-SDR) \cite{sisdr}
as the training objective function, instead of the multi-task loss with
a cross-entropy speaker identification loss used in our previous work of \cite{cd}.
For evaluating the performance, besides the signal-to-distortion ratio (SDR)
of BSSeval \cite{sdr} and SI-SDR, we also report the PESQ \cite{pesq} and STOI \cite{stoi} to measure the speech quality and intelligibility. 

\subsection{Results in SDR/SI-SDR }
\label{subsec:ricd}

Table \ref{tab:cd} shows the performance comparison among different
frameworks in SDR/SI-SDR for female-female (FF), male-male (MM), female-male (FM)
and the whole (Avg) conditions of the reverberated WSJ0-2mix evaluation sets.
The first ten systems, from ``TSB \cite{tsb}" to ``CD-Old-IPD", are all mentioned in our
previous work of \cite{cd}, but their performances are slightly
worse (absolute $<$ 0.2dB SDR) than the numbers we reported in \cite{cd}. Because
they are all only trained using the SI-SDR instead of the multi-task loss.

\begin{table}[t]
  \caption{SDR/SI-SDR (dB) performance of different target speech extraction systems.
  ``SA" represents the speaker adaptation is also performed on the parallel encoder output.
  ``FF", ``MM", ``FM" and ``Avg" represent the female-female, male-male, female-male and average conditions. All the systems are
  trained only using the SI-SDR loss.}
  \label{tab:cd}
  \footnotesize
	\setlength{\tabcolsep}{0.5mm}
  \centering
  \begin{tabular}{l|c|c|c|c|c}
    \toprule
	System  & SA & FF & MM & FM & Avg\\
    \midrule
	 TSB \cite{tsb}    & - & 9.13 / - & 9.47 / - & 12.77 / - & 11.17 / - \\
	 TSB-IPD \cite{tsb}    & - & 10.17 /- & 10.30 / - & 12.49 / - & 11.45 / - \\
    \midrule
	 TSB(our)   & -   & 9.43 / 8.84 & 10.02 / 9.52 & 12.54 / 12.06 & 11.26 / 10.76\\
	 TSB-IPD(our) 	& -   & 10.01 / 9.46 & 10.51 / 10.02 &  12.80 / 12.31  & 11.65 / 11.15 \\
	\midrule
    Para-Enc & -   & 10.83 / 10.21 & 11.52 / 10.99 & 13.12 / 12.61  & 12.26 / 11.72 \\
    Para-Enc                & \checkmark   & 11.31 / 10.68 & 11.91 / 11.41 & 13.31 / 12.79  & 12.56 / 12.02 \\
    \midrule
    CD-Old     & -   & 11.16 / 10.53 & 11.67 / 11.10 & 13.20 / 12.69  & 12.41 / 11.85 \\
    CD-Old            & \checkmark   & 11.53 / 10.92 & 12.02 / 11.52 & 13.60 / 13.09  & 12.78 / 12.26 \\
    CD-Old-Tied           & \checkmark   & 11.41 / 10.82 & 11.83 / 11.27 & 13.53 / 13.02  & 12.67 / 12.13 \\
    CD-Old-IPD             & \checkmark   & 11.20 / 10.58 & 11.79 / 11.27 & 13.39 / 12.88  & 12.55 / 12.01 \\
    \midrule
    CD-Unrolled     & -   & 11.41 / 10.78 & 12.10 / 11.60 & 13.64 / 13.14  & 12.81 / 12.28 \\
    CD-Unrolled            & \checkmark   & 11.33 / 10.70 & 12.27 / \textbf{11.77} & 13.78 / \textbf{13.28}  & 12.92 / 12.39 \\
    \midrule
    CD-Cosine     & -   & 11.30 / 10.62 & 12.09 / 11.57 & 13.61 / 13.12  & 12.77 / 12.24 \\
    CD-Cosine            & \checkmark   & 11.53 / 10.89 & 12.08 / 11.56 & 13.57 / 13.06  & 12.78 / 12.25 \\
    \midrule
    CD-Para-a    & \checkmark & \textbf{11.71 / 11.10} & \textbf{12.28} / 11.76 & \textbf{13.80 / 13.28} & \textbf{13.00 / 12.46}\\
    CD-Para-b 	& \checkmark & 11.41 / 10.74 & 12.15 / 11.64 &  13.68 / 13.17  & 12.84 / 12.30 \\
  	\bottomrule
  \end{tabular}
\end{table}

``TSB \cite{tsb}, TSB-IPD \cite{tsb}" are the state-of-the-art TD-SpeakerBeam (TSB) without
and with IPD features. Because the source code of TSB is not publicly available,
we implemented this algorithm in \cite{cdcode} and the reproduced results shown in
``TSB (our), TSB-IPD (our)" are slightly better than the ones in \cite{tsb}.
Instead of using IPD to exploit the multi-channel spatial information,
the parallel encoder (Para-Enc) \cite{para} was proposed to directly sums each mixture
encoded representation of multi-channels as the input of mask estimator,
and it achieved much better results than IPD. Therefore, our study is also
based on this parallel encoder structure, and ``Para-Enc" is taken as the
strongest baseline for all of the proposed CD-based systems.

``CD-Old, CD-Old-Tied" and ``CD-Old-IPD" are the improved parallel encoder-based
TSE systems using our previously proposed CD mechanism without and with parameters sharing
between encoders, CD together with IPD features respectively. We see that
the ``CD-Old" achieves the best results over other two systems. Sharing encoder parameters
in CD does not bring any extra benefits. And incorporating IPD features
degrades the performance a little bit, this may results from the spatial information
mismatch between IPD features and CD output, because IPD is extracted
in the frequency domain, while the CD is performed in the time-domain.

Comparing ``CD-Old" with ``Para-Enc" baseline, we see that our original CD
does improved the average SDR/SI-SDR from 12.26/11.72 to 12.41/11.85.
However, these performance gains are very limited. By introducing the
unrolled and cosine-based CD, significant performance gains are obtained, and the
unrolled way achieves the best result. Absolute 0.55/0.56 dB (from 12.26/11.72 to 12.81/12.28)
average SDR/SI-SDR gains have been obtained by ``CD-Unrolled" over the ``Para-Enc" baseline.
And these gains are consistent for all the same-gender (FF, MM) and cross-gender (FM)
evaluation conditions.

Moreover, besides the improvements of CD mechanism, the proposed additional
scaling speaker adaptation (SA) on the encoder output also improves the
parallel encoder-``Para-Enc" baseline. By comparing the ``Para-Enc" with (perform the
SA on the $\mathbf{W_2}$ ) and without ``SA", we obtain an absolute 0.3dB
average SDR/SI-SDR improvement. Consistent gains are also achieved in the ``CD-Old" system
(as Fig.\ref{fig:overviews} (b)). However, it's interesting to find that when we
generalize the ``SA" on $\mathbf{W_2}$ of the improved CD, ``CD-Unrolled" or ``CD-Cosine",
the performance improvements become very limited, this may due to the fact that
the differential spatial information extracted by the improved CD is too sparse to
characterize the target speaker identity. This differential information is more
useful to enhance the mixture separation. Therefore, by changing the combination
strategy between CD and SA of parallel encoder outputs as in Fig.\ref{fig:cd_sa}
(a) and (b), we obtain two corresponding results in system ``CD-Para-a" and ``CD-Para-b".
Compared with ``CD-Unrolled (w/o SA)", ``CD-Para-a" still can achieve a further
absolute 0.3 dB SDR/SI-SDR improvement (from 11.41/10.78 to 11.71/11.10) on the ``FF"
evaluation condition, although the improvements on other conditions are relatively
small. That's to say, incorporating the improved CD and SA in an appropriate way
can maximize their individual advantage for improving a TSE system.
Compared with the ``Para-Enc" baseline, the best proposed system ``CD-Para-a"
achieves $8.12/8.71\%$, $6.59/7.00\%$, $5.18/5.31\%$, and $6.03/6.31\%$ relative improvements in SDR/SI-SDR for female-female, male-male, female-male, and average conditions respectively.
It indicates that the improved CD with SA can significantly enhance
the target speech extraction, especially for the same-gender mixtures.

\subsection{Results in PESQ/STOI }
\label{subsec:rps}

\begin{table}[h]
  \caption{PESQ/STOI performance of different target speech extraction systems.
  ``SA" represents the speaker adaptation is also performed on the parallel encoder output.}
  \label{tab:pesq}
  \centering
  \begin{tabular}{l|c|c|c}
    \toprule
	System  & SA & PESQ & STOI\\
    \midrule
	Mixture & - & 1.964 & 0.742 \\
    \midrule
  Para-Enc & - & 2.987 & 0.910 \\
  Para-Enc & \checkmark & 3.047 & 0.915 \\
    \midrule
  CD-Unrolled & - & 3.107 & 0.917 \\
  CD-Unrolled & \checkmark & 3.143 & 0.918 \\
    \midrule
  CD-Para-a & \checkmark & 3.142 & 0.920\\

  	\bottomrule
  \end{tabular}
\end{table}


Table \ref{tab:pesq} shows the PESQ and STOI performance comparison 
among different systems. Compared with ``Para-Enc (w/o SA)", the 
``CD-Unrolled (w/o SA)" achieves $0.120$ and $0.007$ improvements in 
PESQ and STOI respectively. Furthermore, we also see that the additional 
speaker adaptation performed on the parallel encoder output 
is also beneficial for improving the speech quality and intelligibility.

\section{Conclusion}


In this study, two methods are proposed to improve our original 
channel decorrelation algorithm. By expanding the dynamic range of 
decorrelation distribution, the improved algorithm can capture 
better fine-grained inter-channel differential information.
Furthermore, new combination strategies of the improved  
decorrelation and additional target speaker adaptation of 
encoder outputs are also investigated. Extensive experimental results 
demonstrate that both the improvements on the original channel decorrelation 
and target speaker adaptation are very effective to build a 
much better target speech extraction system.
Our future work will focus on generalizing the proposed 
algorithms to other related tasks.

\bibliographystyle{IEEEtran}
\bibliography{mybib}

\begin{thebibliography}{10}
\providecommand{\url}[1]{#1}
\csname url@samestyle\endcsname
\providecommand{\newblock}{\relax}
\providecommand{\bibinfo}[2]{#2}
\providecommand{\BIBentrySTDinterwordspacing}{\spaceskip=0pt\relax}
\providecommand{\BIBentryALTinterwordstretchfactor}{4}
\providecommand{\BIBentryALTinterwordspacing}{\spaceskip=\fontdimen2\font plus
\BIBentryALTinterwordstretchfactor\fontdimen3\font minus
  \fontdimen4\font\relax}
\providecommand{\BIBforeignlanguage}[2]{{%
\expandafter\ifx\csname l@#1\endcsname\relax
\typeout{** WARNING: IEEEtran.bst: No hyphenation pattern has been}%
\typeout{** loaded for the language `#1'. Using the pattern for}%
\typeout{** the default language instead.}%
\else
\language=\csname l@#1\endcsname
\fi
#2}}
\providecommand{\BIBdecl}{\relax}
\BIBdecl

\bibitem{tse}
M.~Delcroix, K.~Zmolikova, K.~Kinoshita, A.~Ogawa, and T.~Nakatani, ``Single
  channel target speaker extraction and recognition with speaker beam,'' in
  \emph{Proc. ICASSP}.\hskip 1em plus 0.5em minus 0.4em\relax IEEE, 2018, pp.
  5554--5558.

\bibitem{tse_related1}
J.~Wang, J.~Chen, D.~Su, L.~Chen, M.~Yu, Y.~Qian, and D.~Yu, ``Deep extractor
  network for target speaker recovery from single channel speech mixtures,'' in
  \emph{Proc. Interspeech}, 2018, pp. 307--311.

\bibitem{tse_related2}
Q.~Wang, H.~Muckenhirn, K.~Wilson, P.~Sridhar, Z.~Wu, J.~R. Hershey, R.~A.
  Saurous, R.~J. Weiss, Y.~Jia, and I.~L. Moreno, ``Voicefilter: Targeted voice
  separation by speaker-conditioned spectrogram masking,'' in \emph{Proc.
  Interspeech}, 2019, pp. 2728--2732.

\bibitem{tse_related3}
C.~Xu, W.~Rao, E.~S. Chng, and H.~Li, ``Spex: Multi-scale time domain speaker
  extraction network,'' \emph{IEEE/ACM Transactions on Audio, Speech, and
  Language Processing}, vol.~PP, no.~99, pp. 1--1, 2020.

\bibitem{tse_related4}
Y.~Hao, J.~Xu, J.~Shi, P.~Zhang, L.~Qin, and B.~Xu, ``A unified framework for
  low-latency speaker extraction in cocktail party environments,'' in
  \emph{Proc. Interspeech}, 2020, pp. 1431--1435.

\bibitem{tse_related5}
J.~Zhao, S.~Gao, and T.~Shinozaki, ``Time-domain target-speaker speech
  separation with waveform-based speaker embedding,'' in \emph{Proc.
  Interspeech}, 2020, pp. 1436--1440.

\bibitem{direct}
G.~Li, S.~Liang, S.~Nie, W.~Liu, M.~Yu, L.~Chen, S.~Peng, and C.~Li,
  ``Direction-aware speaker beam for multi-channel speaker extraction.'' in
  \emph{Proc. Interspeech}, 2019, pp. 2713--2717.

\bibitem{neural}
R.~Gu, L.~Chen, S.-X. Zhang, J.~Zheng, Y.~Xu, M.~Yu, D.~Su, Y.~Zou, and D.~Yu,
  ``Neural spatial filter: Target speaker speech separation assisted with
  directional information.'' in \emph{Proc. Interspeech}, 2019, pp. 4290--4294.

\bibitem{tsb}
M.~Delcroix, T.~Ochiai, K.~Zmolikova, K.~Kinoshita, N.~Tawara, T.~Nakatani, and
  S.~Araki, ``Improving speaker discrimination of target speech extraction with
  time-domain speakerbeam,'' in \emph{Proc. ICASSP}.\hskip 1em plus 0.5em minus
  0.4em\relax IEEE, 2020, pp. 691--695.

\bibitem{ipd}
Z.~Chen, X.~Xiao, T.~Yoshioka, H.~Erdogan, J.~Li, and Y.~Gong, ``Multi-channel
  overlapped speech recognition with location guided speech extraction
  network,'' in \emph{Proc. SLT}.\hskip 1em plus 0.5em minus 0.4em\relax IEEE,
  2018, pp. 558--565.

\bibitem{ss}
F.~Bahmaninezhad, J.~Wu, R.~Gu, S.-X. Zhang, Y.~Xu, M.~Yu, and D.~Yu, ``A
  comprehensive study of speech separation: Spectrogram vs waveform
  separation,'' in \emph{Proc. Interspeech}, 2019, pp. 4574--4578.

\bibitem{ncc}
Y.~Luo, C.~Han, N.~Mesgarani, E.~Ceolini, and S.-C. Liu, ``Fasnet: Low-latency
  adaptive beamforming for multi-microphone audio processing,'' in \emph{Proc.
  ASRU}.\hskip 1em plus 0.5em minus 0.4em\relax IEEE, 2019, pp. 260--267.

\bibitem{tac}
Y.~Luo, Z.~Chen, N.~Mesgarani, and T.~Yoshioka, ``End-to-end microphone
  permutation and number invariant multi-channel speech separation,'' in
  \emph{Proc. ICASSP}.\hskip 1em plus 0.5em minus 0.4em\relax IEEE, 2020, pp.
  6394--6398.

\bibitem{icd}
R.~Gu, S.-X. Zhang, L.~Chen, Y.~Xu, M.~Yu, D.~Su, Y.~Zou, and D.~Yu,
  ``Enhancing end-to-end multi-channel speech separation via spatial feature
  learning,'' in \emph{Proc. ICASSP}.\hskip 1em plus 0.5em minus 0.4em\relax
  IEEE, 2020, pp. 7319--7323.

\bibitem{cd}
J.~Han, X.~Zhou, Y.~Long, and Y.~Li, ``Multi-channel target speech extraction
  with channel decorrelation and target speaker adaptation,'' in \emph{Proc.
  ICASSP}.\hskip 1em plus 0.5em minus 0.4em\relax IEEE, 2021, pp. 6094--6098.

\bibitem{para}
R.~Gu, J.~Wu, S.-X. Zhang, L.~Chen, Y.~Xu, M.~Yu, D.~Su, Y.~Zou, and D.~Yu,
  ``End-to-end multi-channel speech separation,'' \emph{arXiv preprint
  arXiv:1905.06286}, 2019.

\bibitem{tasnet}
Y.~Luo and N.~Mesgarani, ``Conv-tasnet: Surpassing ideal time--frequency
  magnitude masking for speech separation,'' \emph{IEEE/ACM transactions on
  audio, speech, and language processing}, vol.~27, no.~8, pp. 1256--1266,
  2019.

\bibitem{tasnet_relate1}
Z.~Shi, H.~Lin, L.~Liu, R.~Liu, J.~Han, and A.~Shi, ``Deep attention gated
  dilated temporal convolutional networks with intra-parallel convolutional
  modules for end-to-end monaural speech separation.'' in \emph{Proc.
  Interspeech}, 2019, pp. 3183--3187.

\bibitem{tasnet_relate2}
S.~Sonning, C.~Sch{\"u}ldt, H.~Erdogan, and S.~Wisdom, ``Performance study of a
  convolutional time-domain audio separation network for real-time speech
  denoising,'' in \emph{Proc. ICASSP}.\hskip 1em plus 0.5em minus 0.4em\relax
  IEEE, 2020, pp. 831--835.

\bibitem{tasnet_relate3}
J.~Shi, J.~Xu, Y.~Fujita, S.~Watanabe, and B.~Xu, ``Speaker-conditional chain
  model for speech separation and extraction,'' in \emph{Proc. Interspeech},
  2020, pp. 2707--2711.

\bibitem{tasnet_relate4}
Y.~Luo, Z.~Chen, and T.~Yoshioka, ``Dual-path rnn: efficient long sequence
  modeling for time-domain single-channel speech separation,'' in \emph{Proc.
  ICASSP}.\hskip 1em plus 0.5em minus 0.4em\relax IEEE, 2020, pp. 46--50.

\bibitem{tcn}
C.~Lea, R.~Vidal, A.~Reiter, and G.~D. Hager, ``Temporal convolutional
  networks: A unified approach to action segmentation,'' in \emph{European
  Conference on Computer Vision}.\hskip 1em plus 0.5em minus 0.4em\relax
  Springer, 2016, pp. 47--54.

\bibitem{sa}
M.~Delcroix, K.~Zmolikova, T.~Ochiai, K.~Kinoshita, S.~Araki, and T.~Nakatani,
  ``Compact network for speakerbeam target speaker extraction,'' in \emph{Proc.
  ICASSP}.\hskip 1em plus 0.5em minus 0.4em\relax IEEE, 2019, pp. 6965--6969.

\bibitem{wsj0}
Z.-Q. Wang, J.~Le~Roux, and J.~R. Hershey, ``Multi-channel deep clustering:
  Discriminative spectral and spatial embeddings for speaker-independent speech
  separation,'' in \emph{Proc. ICASSP}.\hskip 1em plus 0.5em minus 0.4em\relax
  IEEE, 2018, pp. 1--5.

\bibitem{xu}
\url{https://github.com/xuchenglin28/speaker_extraction/tree/master/simulation}.

\bibitem{cdcode}
\url{https://github.com/jyhan03/channel-decorrelation}.

\bibitem{sisdr}
J.~Le~Roux, S.~Wisdom, H.~Erdogan, and J.~R. Hershey, ``Sdr--half-baked or well
  done?'' in \emph{Proc. ICASSP}.\hskip 1em plus 0.5em minus 0.4em\relax IEEE,
  2019, pp. 626--630.

\bibitem{sdr}
E.~Vincent, R.~Gribonval, and C.~F{\'e}votte, ``Performance measurement in
  blind audio source separation,'' \emph{IEEE transactions on audio, speech,
  and language processing}, vol.~14, no.~4, pp. 1462--1469, 2006.

\bibitem{pesq}
A.~W. Rix, J.~G. Beerends, M.~P. Hollier, and A.~P. Hekstra, ``Perceptual
  evaluation of speech quality (pesq)-a new method for speech quality
  assessment of telephone networks and codecs,'' in \emph{Proc. ICASSP}.\hskip
  1em plus 0.5em minus 0.4em\relax IEEE, 2001, pp. 749--752.

\bibitem{stoi}
C.~H. Taal, R.~C. Hendriks, R.~Heusdens, and J.~Jensen, ``A short-time
  objective intelligibility measure for time-frequency weighted noisy speech,''
  in \emph{Proc. ICASSP}.\hskip 1em plus 0.5em minus 0.4em\relax IEEE, 2010,
  pp. 4214--4217.

\end{thebibliography}

\end{document}